\documentclass[aps,showpacs,preprint]{revtex4}
\usepackage{hyperref}
\usepackage{graphicx}
\usepackage{color}

\usepackage{amsmath}

\def\NOT(#1,#2){\OneQubitGate(#1,#2){$X$}}

\begin{document}
\title{Experimental implementation of encoded logical qubit operations in a perfect
quantum error correcting code}
\author{
Jingfu Zhang$^{1}$, Raymond Laflamme$^{2,3}$, and Dieter Suter$^{1}$\\
$^{1}$Fakult\"{a}t Physik, Technische Universit\"{a}t Dortmund,\\
 D-44221 Dortmund, Germany \\
\it {$^2$Institute for Quantum Computing and Department of
Physics,
University of Waterloo, Waterloo, Ontario, Canada N2L 3G1\\
$^3$Perimeter Institute for Theoretical Physics, Waterloo,
Ontario, N2J 2W9, Canada\\}
 }

\date{\today}

\begin{abstract}
Large-scale universal quantum computing requires the
implementation of quantum error correction (QEC). While the
implementation of QEC has already been demonstrated for quantum
memories, reliable quantum computing requires also the application
of nontrivial logical gate operations to the encoded qubits. Here,
we present examples of such operations by implementing, in
addition to the identity operation, the NOT and the Hadamard gate
to a logical qubit encoded in a five qubit system that allows
correction of arbitrary single qubit errors. We perform quantum
process tomography of the encoded gate operations, demonstrate the
successful correction of all possible single qubit errors and
measure the fidelity of the encoded logical gate operations.
\end{abstract}
\pacs{03.67.Pp,03.67.Lx} \maketitle

{\it Introduction.--}Quantum computers can solve certain problems
exponentially faster than classical computers
\cite{NielsenChuang,StolzeSuter}. An essential precondition for
realizing this potential is the preservation of the coherence
between quantum states. This requirement makes the implementation
of quantum computing much more challenging than for classical
devices. Reliable quantum computing \cite{3921} is possible, in
principle, provided quantum error correction (QEC) schemes can be
implemented with a fidelity above a certain threshold
\cite{NielsenChuang,Knill05,Got06,gottesman97,accuQIC}. Every QEC
code has an overhead in terms of gate operations and additional
(ancilla) qubits. The protection of a single qubit against
arbitrary single-qubit errors requires at least five physical
qubits \cite{RayPRL96,BDS+96}.

\begin{figure}
\includegraphics[width=3in]{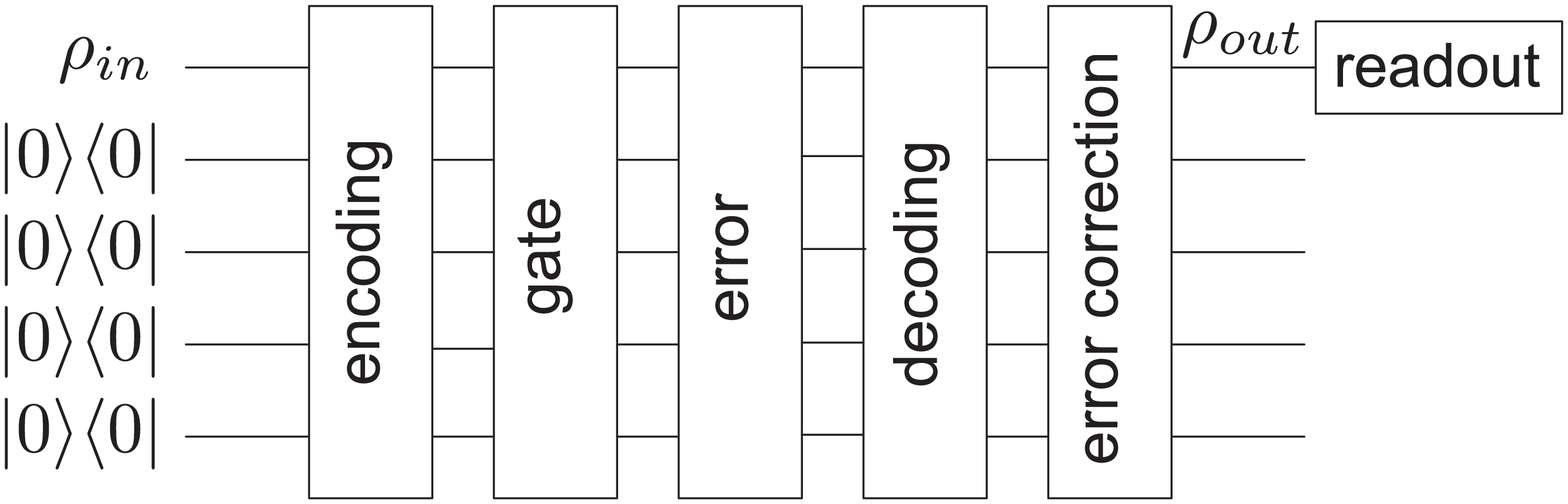} 
\caption{Outline of the quantum algorithm.} \label{figoutline}
\end{figure}

Over the last few years, a series of experiments were performed
that demonstrated that QEC is indeed capable of protecting quantum
states against generated (artificial) errors or decoherence
induced by the environment (see, e.g.
\cite{PhysRevLett.81.2152,RayPRL01,PhysRevLett.107.160501,nature.432.602,nature.482.382}).
However, the realization of reliable quantum computing requires
more than the preservation of information: it must also be
possible to process the encoded information by applying logical
gate operations to the protected qubits. Here, we present an
experimental demonstration of such gate operations on an encoded
qubit. For encoding, we use a five bit QEC code
\cite{RayPRL96,BDS+96} that allows correction of arbitrary
single-qubit errors (a so-called perfect QEC code) and demonstrate
successful gate implementation and error correction. The result is
a fault-tolerant implementation of the corresponding gate
operations. Fig. \ref{figoutline} summarizes the scheme: it starts
with encoding the input state of the first qubit into five
physical qubits. To the resulting state, we apply one of three
single-qubit gates - the identity, NOT or the Hadamard gate. As
the third step, we apply another operation, which is either the
identity (corresponding to no error) or one of the fifteen
possible single-qubit errors. In the fourth step, the information
is decoded, i.e. the output is extracted into the state of the
first physical qubit. In the fifth and final step, possible errors
are detected and corrected.

For the physical qubits, we use a system of five nuclear spins.
The molecule containing the spins is dissolved in an anisotropic
solvent. The resulting magnetic dipole couplings between the
nuclear spins are significantly stronger than the more frequently
used scalar couplings and therefore result in a speedup of the
gate operations by approximately an order of magnitude. The
complete operation can therefore be completed within a time period
significantly shorter than the coherence time of the system.

\begin{figure}
\includegraphics[width=2.5in]{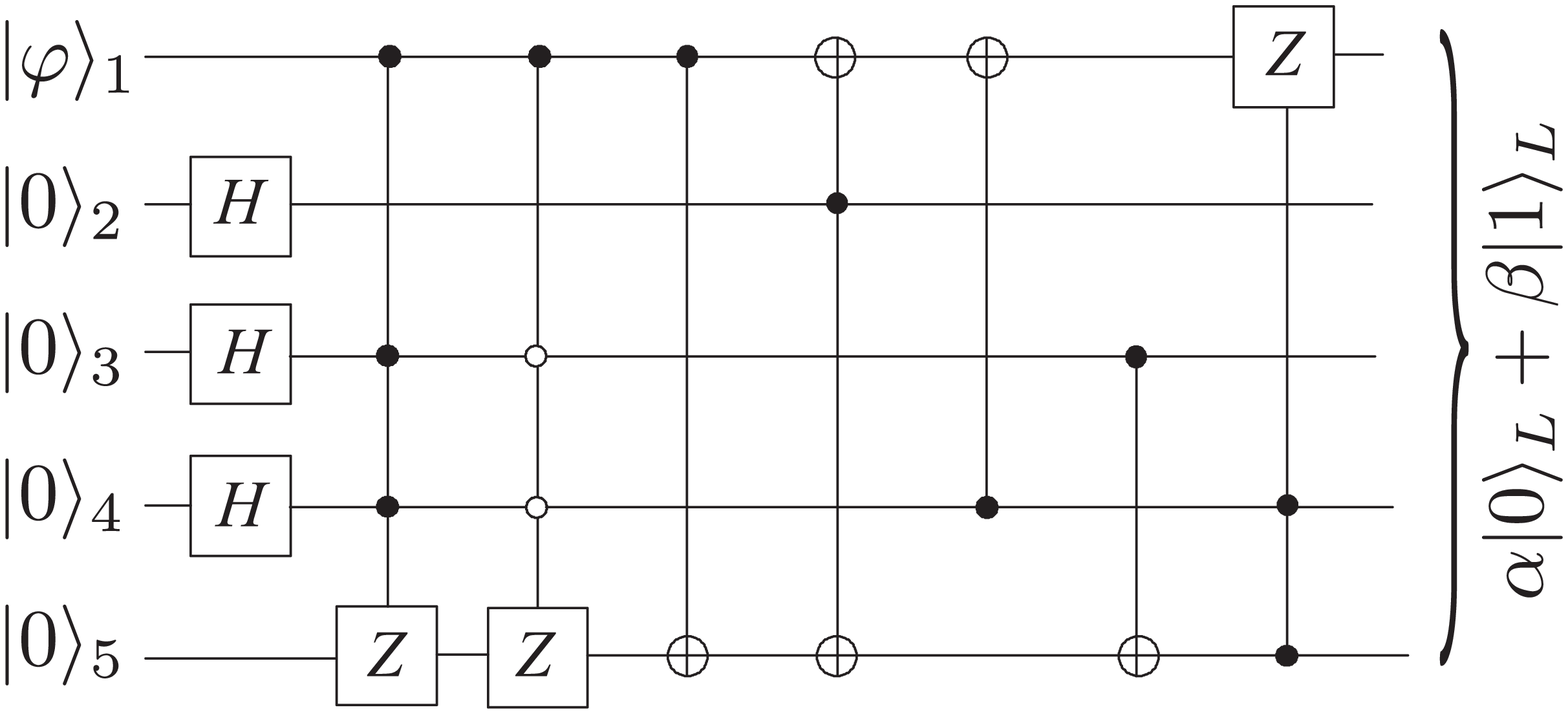} 
\caption{Network representation of the encoding operation.
The input state $|\varphi\rangle = \alpha|0\rangle+\beta|1\rangle$ is initially
encoded in qubit $1$.
Qubits 2--5 are the syndrome bits.
$H$ denotes a Hadamard transform, and $Z$
the $\pi$ phase gate. The box or $\bigoplus$ with a connected
vertical line indicate controlled gates, where the filled or
empty circle marks the control qubit. The operations are
conditional on the control qubit being in state $|1\rangle$
(filled circle) or $|0\rangle$ (empty circle), respectively. The
output of the circuit is the five-qubit state $\alpha|0\rangle_{L}+
\beta|1\rangle_{L}$.} \label{figcir}
\end{figure}

{\it Five-qubit  error correcting code.--}
Fig. \ref{figcir}  shows the quantum circuit for the five-qubit QEC code \cite{RayPRL96}.
Qubit $1$ is the register bit
carrying the input state $|\varphi\rangle =
\alpha|0\rangle+\beta|1\rangle$, where $\alpha$ and $\beta$ are
arbitrary complex numbers with $|\alpha|^2+|\beta|^2 = 1$. The
other four qubits are initialized in the state $|0\rangle^{\otimes
4}$.
The unitary operation $U_{en}$ implemented by the circuit
encodes the state $|\varphi\rangle$ into a logical state as
\begin{equation}\label{logicalarb}
 U_{en}(\alpha|0\rangle+\beta|1\rangle)|0000\rangle = \alpha|0\rangle_{L}+\beta|1\rangle_{L},
\end{equation}
where
\begin{eqnarray}\label{Logical0}
|0\rangle_{L} &\equiv&
\frac{1}{\sqrt{8}}(|00000\rangle-|10111\rangle-|01011\rangle+|11100\rangle
\nonumber \\
   &+&|10010\rangle
   +|00101\rangle+|11001\rangle+|01110\rangle)
 \end{eqnarray}
\begin{eqnarray}\label{Logical1}
|1\rangle_{L} &\equiv& \frac{1}{\sqrt{8}}(|11111\rangle-
|01000\rangle+ |10100\rangle- |00011\rangle \nonumber \\
    &+& |01101\rangle + |11010\rangle - |00110\rangle -
    |10001\rangle)
 \end{eqnarray}
are the computational basis states of the logical qubit.
The decoding operation $U_{de}$ is
the inverse operation of $U_{en}$, i.e., $U_{de} = U_{en}^{\dag}$.

The five-qubit QEC code can detect and correct arbitrary
single-qubit errors. The possible single-qubit errors for a
five-qubit system can be written as bit flip errors B$k$, phase
flip errors S$k$ and combined bit- and phase flip errors BS$k$,
where $k = 1 \dots 5$ indicates the affected qubit. These fifteen
errors, together with the identity operation $E$ define the
possible outcomes if only single-qubit errors occur. The $2^4$
possible states of the four syndrome qubits can distinguish
between these 16 different outcomes. This is used by the error
correction step, which is a unitary operation on the first qubit,
controlled by all four syndrome qubits. Without the encoded gate
operations, this code was implemented previously in a system of
weakly coupled spin qubits \cite{RayPRL01}.

\begin{figure}
\includegraphics[width=3in]{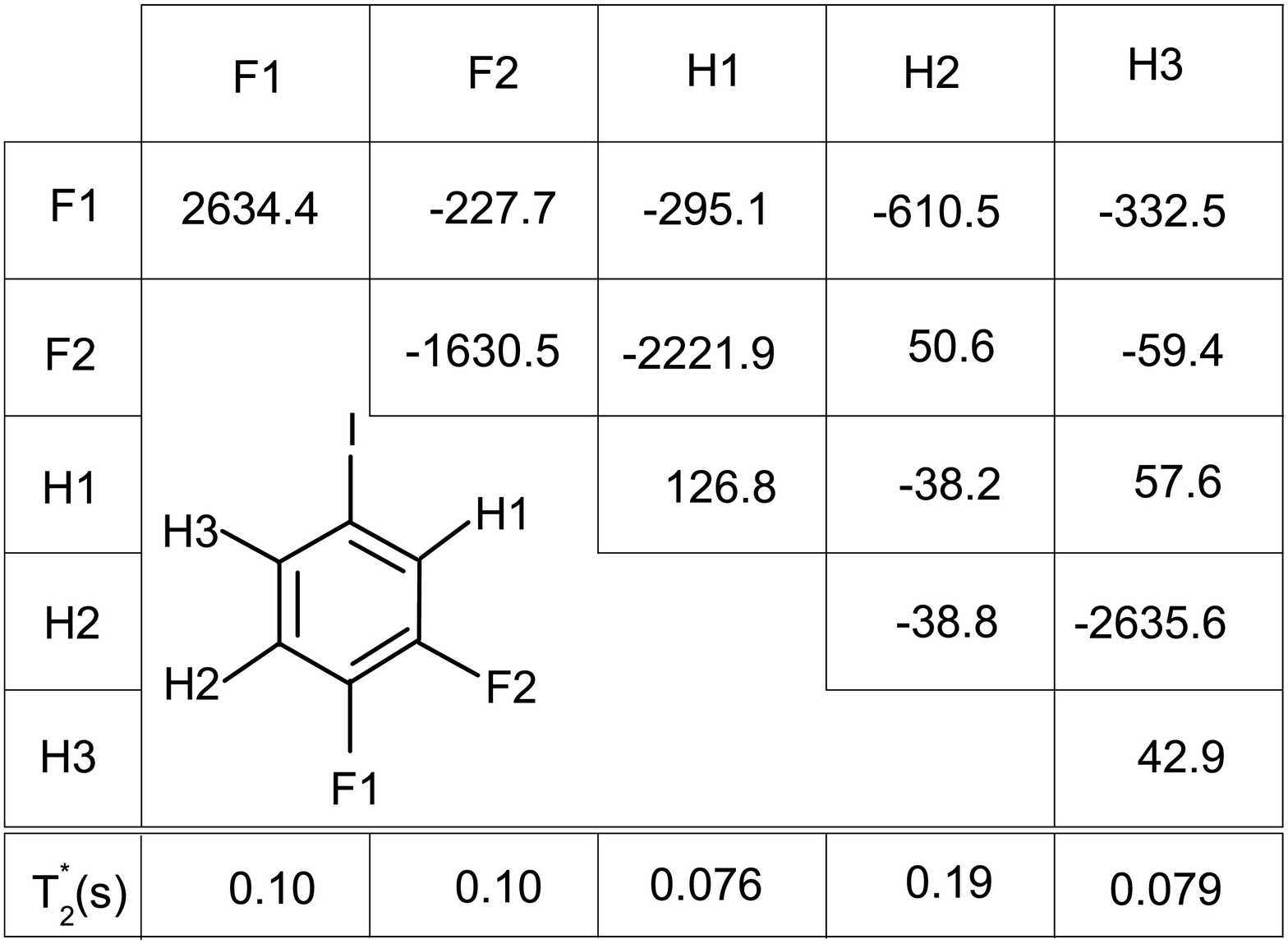} 
\caption{Parameters of the spin qubits in the molecule
1,2-difluoro-4-iodobenzene. The inset shows the structure, where
the five qubits 1--5 are spins F1, F2, H1, H2, and H3,
respectively. The chemical shifts with respect to the transmitter
frequencies of the proton and fluorine spins are shown as the
diagonal terms and the dipolar couplings between spins are shown
as the off-diagonal terms in units of Hz.
The effective relaxation times $T^{*}_2$ are determined by
fitting the peaks in the spectra. 
}
\label{figsch}
\end{figure}

 {\it Experimental protocol.--}For the experimental implementation,
 we used the two fluorine and three proton spins of
the molecule of 1,2-difluoro-4-iodobenzene,  whose structure is
shown in Fig. \ref{figsch}. The molecule was dissolved in the
liquid-crystal solvent ZLI-1132 to retain the dipolar couplings
between the spins. We denote the two fluorine spins  F1 and F2 and
the three protons H1, H2, and H3 as qubits 1--5. Data were taken
with a Bruker Avance II 500 MHz spectrometer. The relevant
Hamiltonian of the dipolar coupled spin system is, in frequency
units,
   $\mathcal{H} = \sum_{i} \mathcal{H}^{c}_{i} + \sum_{i<j} \mathcal{H}^{d}_{ij}$,
where $\mathcal{H}_{i}^{c} =\pi \nu_{i}Z_{i}$ is the Zeeman
Hamiltonian and  $\mathcal{H}_{ij}^{d} = \pi D_{ij}Z_{i}Z_{j}$
describes the dipolar coupling Hamiltonian for the heteronuclear
spins $i$ and $j$, or $\mathcal{H}_{ij}^{d} = (\pi
D_{ij}/2)(2Z_{i}Z_{j}-X_{i}X_{j}-Y_{i}Y_{j})$ for the homonuclear
case. The indices $i$ and $j$ run over all five spins, $X_{i}$,
$Y_{i}$, $Z_{i}$ denote the Pauli matrices, $\nu_{i}$  the
chemical shift of spin $i$, and $D_{ij}$  the coupling constant.
 Compared to the dipolar couplings, the scalar
couplings between the spins are negligibly small or can be merged
in the dipolar couplings \cite{SM}. To determine the numerical
values of the Hamiltonian parameters, we measured different
experimental spectra, using multiple quantum NMR \cite{HOCNMR} and
heteronuclear decoupling. These spectra were used in a fitting
process which yielded the parameters  listed in Fig. \ref{figsch}.

We prepared the initial states $X\mathbf{0000}$ or
$Y\mathbf{0000}$, with $\mathbf{0}\equiv|0\rangle\langle0|$ by the
circuit in Ref. \cite{knill}. The required unitary operations were
implemented by two optimized shaped pulses designed with a GRAPE
algorithm \cite{grape,Ryan2008,SM}. The experimental spectrum of
the state $X\mathbf{0000}$ is illustrated in the Supplementary
Material \cite{SM}. It contains a single main peak, which is the
signature of a pseudopure state (PPS) \cite{knill}. By comparing
with a reference spectrum obtained from the thermal state, we
estimate that the polarization of this state is $\approx 0.72$ of
the maximum polarization that could result from an ideal
preparation. For the following experiments, we normalized the
spectra to this one, so the fidelities of the QEC protocol do not
include losses during the PPS preparation
 \cite{SM}. 
 The $Z\mathbf{0000}$ state was
prepared by applying an additional $\pi/2$ rotation to the
$X\mathbf{0000}$ state. After the encoding step, we applied one of
the following gate operations to the logical qubit: the identity
$E$, the NOT gate (up to a known phase), or the  Hadamard gate.

The NOT gate $N_L$ for the logical qubit is relatively simple to implement for this code,
since it is transversal, i.e. it can be written as
\begin{equation}\label{NotL}
  N_L = R^{y}(\pi)^{\otimes 5},
\end{equation}
where $ R^{y}(\pi) =e^{-i\pi Y /2}$.
This relation can be verified by comparing the definitions of the logical states in Eqs.
(\ref{Logical0}--\ref{Logical1}).
The elements of $N_L$ in the logical basis are represented as $\langle 0_L|N_L|0_L\rangle =
\langle 1_L|N_L|1_L\rangle =0$, $\langle 1_L|N_L|0_L\rangle = i$,
and $\langle 0_L|N_L|1_L\rangle = -i$.

In contrast to the NOT gate, the Hadamard gate is not transversal
in this code \cite{BeiZeng07}. We therefore have to design an
operation that implements this gate in the $2^5$ dimensional
Hilbert space of the five-qubit system. In this space, the
Hadamard gate should generate a $\pi$ rotation around the (1,0,1)
axis of the two-dimensional subspace spanned by the states
$|0\rangle_L$ and  $|1\rangle_L$ and an identity operation on the
other 30 states. The corresponding unitary operator thus has the
matrix representation
\begin{equation}\label{HL32}
 H_L = \left(%
\begin{array}{ccccccc}
  \frac{1}{\sqrt{2}} & \frac{1}{\sqrt{2}}  &   &   &   \\
  \frac{1}{\sqrt{2}} & -\frac{1}{\sqrt{2}} &   &  &   \\
    &   & 1  &   &   \\
    &   &   & \ddots  &   \\
      &   &   &  & 1 \\
\end{array}
\right),
\end{equation}
where zero elements are not shown.
This unitary
operation can again be implemented by an optimized shaped pulse,
designed in the same way as the other pulses described above.

After the unitary gate operation, we also applied B$k$, BS$k$, or
S$k$ errors to the individual qubits, using single-qubit $\pi$
rotations around the $x$, $y$, or $z$-axis.

For a quantitative evaluation of the algorithm's performance, we
used quantum process tomography (QPT) \cite{QPTini}
of the complete algorithm represented in Fig. \ref{figoutline}.
The process can be completely characterized by its $\chi$ matrix
\cite{NielsenChuang}, which maps an arbitrary input state
$\rho_{in}$ into the output state
$\rho_{out}= \sum_{kl}\chi_{kl}e_{k}\rho_{in}e_{l}^{\dag}$.
Here the operators $e_{k,l} \in \{E, X, -iY, Z\}$ denote the basis
set for describing the process, and the indices $k,l=1$, ..., $4$
run over the elements of the basis set. The measurement of $\chi$
requires the preparation of four input states $\rho_{in} =E$, $X$,
$Y$, and $Z$. For each $\rho_{in}$, we determined the output state
$\rho_{out}$ by quantum state tomography.  Since the unit operator
$E$ is always time-independent, the corresponding input state is
omitted, assuming the output state is $E$. The quantum state
tomography is performed by measuring in one experiment the
transverse magnetization and in a second experiment applying a
$\pi/2$ readout pulse and then measuring the transverse
magnetization. The measured FIDs were Fourier transformed and the
resulting spectra were fitted to the theoretical spectra by
adjusting as a single parameter the overall amplitude. Having the
output states for the four input states, we determined the the
$\chi$ matrix using the established strategy \cite{NielsenChuang}.
 For each process, we quantified the performance by comparing the
experimental ($\chi_{exp}$) and theoretical ($\chi_{th}$) $\chi$
matrices via the fidelity \cite{chi2}
\begin{equation}\label{Fchi}
 F_{\chi}=|{\rm Tr}(\chi_{exp}\chi_{th}^{\dag})|/\sqrt{{\rm Tr}(\chi_{exp}\chi_{exp}^{\dag}){\rm Tr}(\chi_{th}\chi_{th}^{\dag})}.
\end{equation}

\begin{figure} 
\includegraphics[width=5.5in]{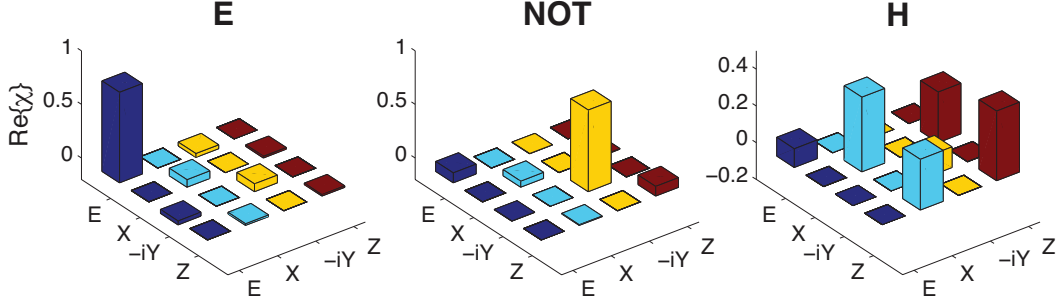} 
\caption{Bar plots of the real parts of the $\chi$ matrices for
the identity, NOT and Hadamard gates in logical states. The RMS
values of the imaginary parts are 0.035, 0.0069, and 0.017,
respectively.} \label{figGchi}
\end{figure}

{\it Experimental results.--}We first checked the encoding,
encoded gate and decoding operations with a simplified
experimental scheme. Compared to the full scheme shown in Fig.
\ref{figoutline}, we omitted the the error and error correction
operations. Fig. \ref{figGchi} shows the experimental results for
the process matrices $\chi$ as bar plots. 
 The corresponding matrices for the ideal
operations have $\chi_{11}=1$ for the identity operation,
$\chi_{33}=1$ for the NOT operation, and
$\chi_{22}=\chi_{44}=\chi_{24}=\chi_{42}=0.5$ for the Hadamard
gate. All other elements should vanish. We find a good qualitative
agreement between the theoretical and experimental values, with
fidelities of 0.979, 0.983, and 0.956, for the identity, NOT and
Hadamard gates, respectively.

\begin{figure}
\includegraphics[width=3in]{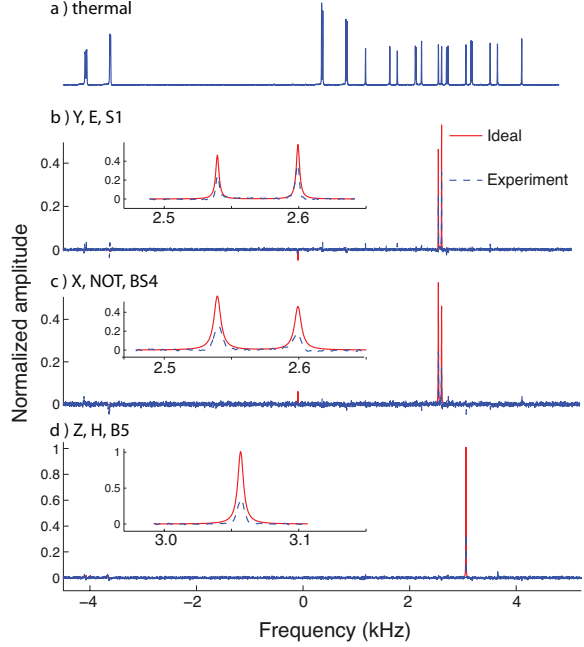} 
\caption{(Color online.) Spectra of the fluorine spins obtained
for different versions of the experiment. a) Reference spectrum
obtained by applying a $\pi/2$ pulse to the thermal equilibrium
state. b) Spectrum of the output state, starting with the initial
state $Y$, encoding, applying the identity operation, S1 error and
error correction. c) Same as b), but $\rho_{in} = X$, NOT gate and
BS4 error. d) Same as b), but $\rho_{in} = Z$, Hadamard gate and
B5 error. The dashed and solid curves indicate the experimental
and simulated spectra, which are normalized to the initial
pseudopure states. The lines with the main signal contributions
are shown enlarged as insets. } \label{figspectra}
\end{figure}

\begin{figure}
\includegraphics[width=5.5in]{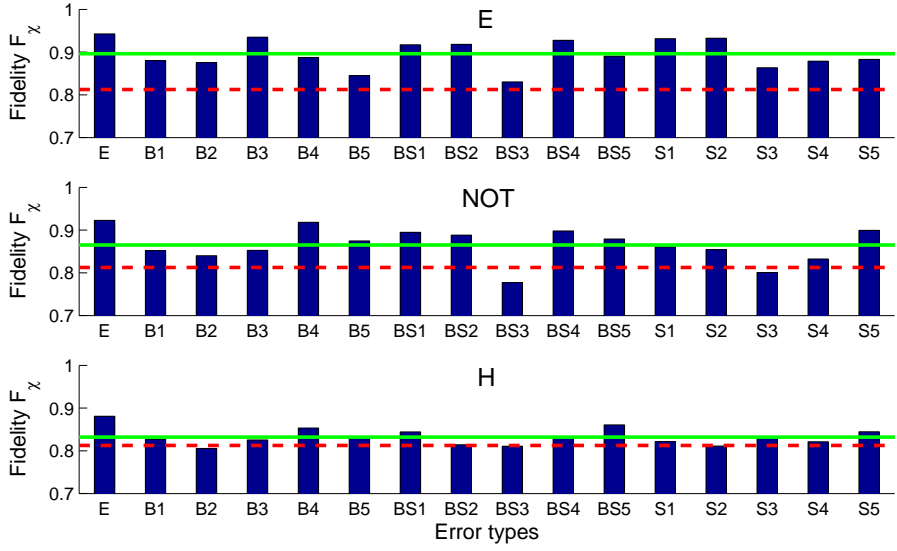} 
\caption{Bar plots showing the experimentally determined
fidelities for the identity, NOT and Hadamard gates applied to the
encoded qubit after applying error operations and error
correction. The average fidelity is shown as the solid line.
Compared to the average fidelity without error correction, shown
as the dashed line, the encoding - decoding - error correction
scheme improves the fidelity. } \label{figFchi}
\end{figure}

We next tested the full implementation, including errors and error
correction. Fig. \ref{figspectra} illustrates the results for
three out of 48 experiments. In the first example (trace b), we
initialized the system to the $Y$ state and applied an $E$ gate
and S$1$ error operation, followed by the decoding and error
correction steps. Trace c) shows the corresponding results for the
$X$ initial state, NOT gate and BS$4$ error and d) for $Z$ initial
state, $H$ gate and B$5$ error. The insets show enlarged partial
spectra containing the main signal components, with the
experimental spectra represented by dashed lines, the ideal
spectra as full lines. Experimental and theoretical curves were
both normalized to the spectra of the initial PPS. Figure
\ref{figFchi} shows the measured fidelities for each of the 48
different experiments as a bar plot. The solid horizontal line
shows the average fidelity for each type of gate, averaged over
the 16 different error conditions.

To assess the usefulness of the scheme, we compare the achieved
fidelities to an idealized experiment where we do not use QEC, but
the same 16 error conditions can occur, with equal probabilities.
In this case, the three single qubit errors acting on the first
qubit result in zero fidelity, while the other 13 error conditions
(the identity and the single-qubit errors on the ancilla qubits)
result in fidelities of one. Averaged
over these 16 reference experiments, we would thus expect an
average fidelity of $13/16 = 0.8125$. This value is shown in the
figure as the horizontal dashed line. The average fidelity for the
three different gates exceeds this reference value by 0.0837,
0.0528 and 0.0196, for the identity, NOT and Hadamard gates,
respectively. This shows that the performance of the QEC scheme is
high enough to compensate the additional errors affecting the
syndrome qubits as well as the errors due to experimental
imperfections of the encoding, decoding and error correction
steps.

{\it Discussion.--}The fidelities achieved in these demonstration
experiments are still lower than the threshold fidelities required
for scalable quantum computation. We have identified four main
causes for the observed reduction of fidelity: (i) Stability and
homogeneity of the magnetic and radio frequency (r.f.) fields were
less than ideal, since the lack of deuterium in the sample did not
allow operation of the lock system of the spectrometer. (ii)
Finite accuracy of the Hamiltonian parameters. (iii) Finite
coherence time: the duration of the experiment is in the range of
$26$--$64$ ms (excluding the preparation of the initial PPS),
which is comparable to the measured
$T_{2}^{*}$. 
(iv)
Deviations between the calculated pulse shapes and those acting on
the spins, which are caused mostly by nonlinearities of the
spectrometer hardware. Our efforts to suppress these imperfections
concentrated on 1) using short gate durations, 2) designing the
pulses robust to variations of the r.f. strength 
and insensitive to frequency offsets 
and 3) `fixing' the pulses by measuring the
actual amplitudes of the control gates with a pick up coil and
adjusting the amplitudes to minimize the difference between
theoretical and experimental values \cite{pulsefixer}.

For a better understanding of the sources of errors, we performed a quantitative
analysis of one of the experiments: the Hadamard gate combined with a
phase flip error at qubit 5 and the corresponding error correction \cite{SM}.
Combining experimental results and simulation, we estimate that the field
inhomogeneity and the imperfect implementation of the pulses
contribute  $\approx 0.09$ to the loss of  fidelity. Additionally
the limit of $T_2$
 and the imprecision in characterizing the Hamiltonian,
contribute $\approx 0.04$ and $0.03$ to the loss of fidelity,
respectively.
  The simulation with $T_2^*$ results in a fidelity below
the experimental value, indicating that the coherence time in the
experiment is longer than $T_2^{*}$. This is expected if the gate
operations refocus some of the inhomogeneity of the system.

{\it Conclusion.--}The results presented here are a first
demonstration of one of the most important preconditions for
reliable quantum computation:  the implementation of logical gate
operations on encoded qubits. While QEC has been demonstrated on
quantum memories, its combination with \emph{processing} of
quantum information is an important milestone towards the
implementation of reliable quantum computing. The present
demonstration used  three single-qubit gate operations acting on a
logical qubit encoded in a perfect five-qubit QEC code. In the
earlier implementation of this code \cite{RayPRL01}, without the
gate operations,  the experimental duration was $>300$ ms.
Compared to that, we have reduced the duration of the experiment
by approximately one order of magnitude by using a dipolar coupled
system, whose interactions are stronger than the scalar couplings
used before. The reduction of the gate operation times opened the
possibility to implement in addition the logical gate operations.
With better homogeneity and more precise Hamiltonian parameters,
it should be possible to improve the experimental fidelity. The
recent theoretical progress in measuring Hamiltonians with dipolar
couplings \cite{Denis} and optimal algorithms in pulse finding
\cite{double} should be helpful for this purpose. This will allow
us to control larger systems, encoding multiple qubits and
implementing multi-qubit gate operations, such as CNOT in the
encoded qubits. If the fidelity can be improved sufficiently to
reach the error threshold, the combination of QEC with logical
gate operations will pave the way to reliable quantum computation.

{\it Acknowledgments.--}The authors acknowledge  A. M. Souza, B.
Zeng, M. Grassl, D. A. Trottier, and V. J. Villanueva for helpful
discussions, and M. Holbach and J. Lambert for help in experiment.
This work is supported by the Alexander von Humboldt Foundation,
and the DFG through Su 192/19-1.


\end{document}